\pdfoutput=1 
\documentclass{JINST}

\title{A Method to Simulate the Observed Surface Properties of Proton Irradiated Silicon Strip Sensors }

\author{T. Peltola$^a$\thanks{Corresponding author.}~\thanks{On behalf of the CMS Tracker collaboration.}~ 
, A. Bhardwaj$^b$, R. Dalal$^b$, R. Eber$^c$, T. Eichhorn$^d$, K. Lalwani$^b$, A. Messineo$^e$, M. Printz$^c$
 and K. Ranjan$^b$\\
\llap{$^a$}Helsinki Institute of Physics (FI)
,\hspace{2 mm}\llap{$^b$}University of Delhi (IN)
,\hspace{2 mm}\llap{$^c$}Karlsruhe Institute of Technology (DE),\hspace{2 mm}\llap{$^d$}Deutsches Elektronen-Synchrotron (DE),
\hspace{2 mm}\llap{$^e$}Universit\'{a} di Pisa \& INFN sez. di Pisa (IT)\\
E-mail: \email{timo.peltola@helsinki.fi}}

\abstract{During the scheduled high luminosity upgrade of the LHC, the world's largest particle physics accelerator at CERN, the position sensitive silicon detectors installed in the vertex and tracking part of the CMS experiment will face a more intense radiation environment than the present system was designed for. To upgrade the tracker to the required performance level, extensive measurements and simulation studies have already been carried out. \\
A defect model of Synopsys Sentaurus TCAD simulation package for the bulk properties of proton irradiated devices has been producing simulations closely matching to measurements of silicon strip detectors. However, the model does not provide the expected behavior due to the fluence increased surface damage. The solution requires an approach that does not affect the accurate bulk properties produced by the proton model, but only adds to it the required radiation induced properties close to the surface. These include the observed position dependency of the strip detector's charge collection efficiency (CCE). \\
In this paper a procedure to find a defect model that reproduces the correct CCE loss, along with other surface properties of a strip detector up to a fluence $1.5\times10^{15}$ 1 MeV $\textrm{n}_{\textrm{\tiny eq}}\textrm{cm}^{-2}$ ($\Phi_{\textrm{\tiny eq}}$), will be presented. When applied to CCE loss measurements at different fluences, this method may provide means for the parametrization of the accumulation of oxide charge at the SiO$_2$/Si interface as a function of dose.} 


\begin{document}
\section{Introduction}\label{sec:1}
\paragraph{} The planned upgrade of the LHC accelerator at CERN, namely the high luminosity (HL) phase of the LHC (HL-LHC scheduled for 2023), will enable the use of maximal physics potential of
the machine. The high integrated luminosity of 3000 fb$^{-1}$ after 10 years of operation, resulting in expected fluence above $1\times10^{15}$ $\textrm{n}_{\textrm{\tiny eq}}$cm$^{-2}$ for strip sensors $\sim$20 cm from vertex, will expose the tracking system at HL-LHC to a radiation environment that is beyond the capability of the present system design. This requires
the upgrade of the all-silicon central trackers that will be equipped with higher granularity as well as radiation hard sensors, which can withstand higher radiation levels and higher occupancies also
in the innermost layers closest to the interaction point. 
For the upgrade, comprehensive measurements and simulation studies for silicon sensors of different designs and materials with sufficient radiation tolerance have been launched.

Complementing measurements, simulations are an integral part of the R$\&$D of novel silicon radiation detector designs with upgraded radiation hardness. By being able to verify experimental results, the numerical simulations will also gain predictive power. This can lead to reduced time and cost budget in detector design and testing.

The simulations in this study were carried out using the Synopsys Sentaurus\footnote{http://www.synopsys.com} finite-element Technology Computer-Aided Design (TCAD) software framework. A radiation defect model that combines experimentally matching surface damage properties with the bulk damage properties of the proton model \cite{bib1} is developed and a comparison with corresponding results produced by the Silvaco Atlas\footnote{http://www.silvaco.com} simulation tool is made.

\section{Surface damage in silicon strip sensors}\label{sec:2}
\paragraph{} At the high radiation environment of the LHC, defects are introduced both in the silicon substrate (bulk damage) and in the SiO$_2$ passivation layer, that affect the sensor performance through the interface with the silicon bulk (surface damage). Bulk damage degrades detector operation by introducing deep acceptor and donor type trap levels \cite{bib2}. Surface damage consists of a positively charged layer accumulated inside the oxide and interface traps may be created close to the interface with silicon bulk \cite{bib3}. These are approximated in the simulation by placing a fixed charge at the interface. High oxide charge densities $Q_{\textrm{\tiny f}}$ are detrimental to the detector performance since the electron layer generated under the SiO$_2$/Si interface can cause very high electric fields near the p$^+$ strips in p-on-n sensors and loss of position resolution in n-on-p sensors by providing a conduction channel between the strips.

Important strip sensor surface characteristics include the interstrip capacitance $C_{\textrm{\tiny int}}$, interstrip
resistance $R_{\textrm{\tiny int}}$, position dependency of the charge collection efficiency CCE($x$) and electric field distribution between strips $E$($x$). $C_{\textrm{\tiny int}}$ and $R_{\textrm{\tiny int}}$ are defined as the capacitance or resistance of an individual strip to its adjacent neighbours and they contribute to strip noise and strip isolation, respectively. Understanding the dependencies of CCE($x$) will lead to more complete interpretation of the degradation of the CCE with fluence. High $E$($x$) can induce detector breakdown or avalanches that can result in non-Gaussian noise events. 

A radiation damage model tuned from the PTI-model \cite{bib4} for proton irradiation at $T=\textrm{-}20^{\circ}\textrm{C}$ is able to reproduce accurately the effects of bulk damage in silicon for fluence range $10^{14}$ $\sim$ $1.5\times10^{15}$ $\textrm{n}_{\textrm{\tiny eq}}$cm$^{-2}$ \cite{bib1}. The proton model does not reproduce the measured $R_{\textrm{\tiny int}}$, $C_{\textrm{\tiny int}}$ and CCE($x$), for expected high accumulated surface charges $Q_{\textrm{\tiny f}}$ due to strip shortening. Hence, the following work has been done to overcome this shortcoming. \\

\section{Simulated device structures}\label{sec:3}
\paragraph{} Within the HPK campaign \cite{bib5} two main device structures have been produced and hence simulated: diodes and segmented sensors. 
To enable comparison with measurements, the structures have been simulated following the real device features as close as possible.

\begin{figure}[tbp] 
\centering 
\includegraphics[width=.3\textwidth]{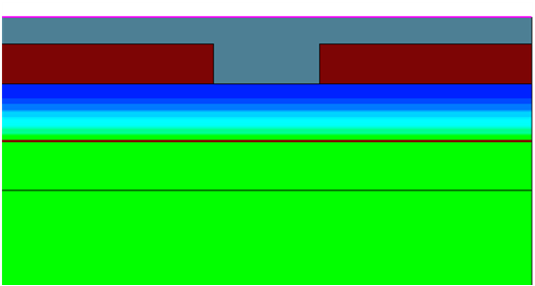}
\includegraphics[width=.56\textwidth]{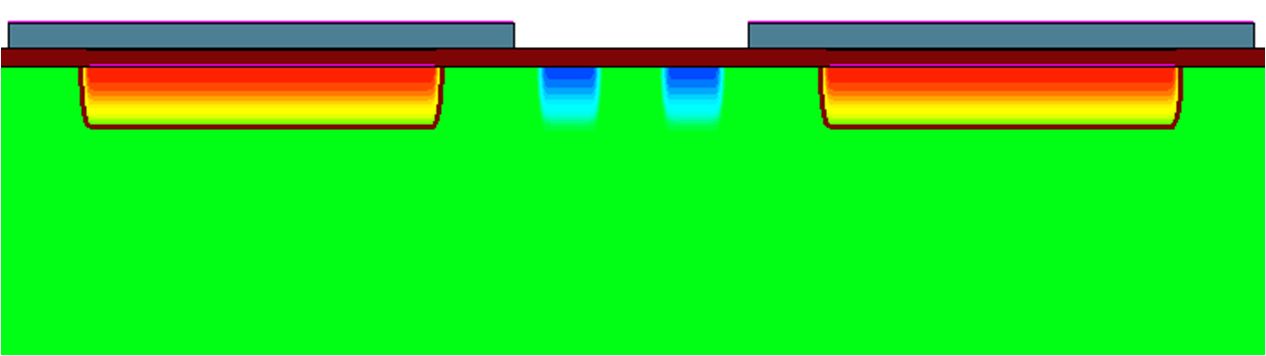}
\caption{Simulated structures. (left) DC-coupled p-on-n diode with via-structure and (right) part of the n-on-p  5-strip sensor front surface with double p-stop strip isolation structure (not to scale). Colors: p$^+$/n$^+$ doping (blue/red), bulk doping (green), oxide (brown) and Al (gray). The aluminum layer is repeated at the backplane where the reverse bias voltage is applied.}
\label{fig:1}
\end{figure}

A diode structure was used to find a matching transient current shape between the defect model under development and the proton model. This was done for simplicity and minimized simulation time since at this point only bulk properties were under investigation, as will be presented in the following sections. Presented on the left side of figure~\ref{fig:1} is the front surface of the simulated $1\times10\times300$ $\mu\textrm{m}^3$ p-on-n diode. The thickness of the oxide on the front surface was 750 nm and the Al layers on the front and backplanes were 500 nm thick. Bulk doping was $3.4\times10^{12}$ $\textrm{cm}^{-3}$ and the heavily doped boron and phosphorus implantations on the front and backplane, respectively, had the peak concentrations $5\times10^{18}$ $\textrm{cm}^{-3}$ with decay to the bulk doping level within 1.0 $\mu\textrm{m}$ depth using error function. 

For the actual surface characteristics simulations a 5-strip structure was chosen to avoid any non-uniformities from border effects on the mesh formation at the center part of the device. Pictured on the right side of figure~\ref{fig:1} are the second and centermost strips of a 5-strip n-on-p sensor with isolation between strips provided by a double p-stop implantation. The strip sensor configuration is region number 5 from the MSSD structure used in the HPK campaign, i.e. pitch is 120 $\mu\textrm{m}$, strip width is 41 $\mu\textrm{m}$, implant width is 28 $\mu\textrm{m}$ and the strip length is 3.049 cm. The p-stop widths are 4 $\mu\textrm{m}$ and the spacing between p-stops is 6 $\mu\textrm{m}$, while peak doping of $1\times10^{16}$ $\textrm{cm}^{-3}$ decays to the bulk doping level within 1.5 $\mu\textrm{m}$, having equal depth with n$^+$ strip implants. The aluminum thicknesses and peak doping values for bulk and implants were as within the diode, while the front surface oxide thickness was set to 250 nm. For the 200 $\mu\textrm{m}$ active thickness n-on-p sensor (200P) the total bulk thickness was 320 $\mu\textrm{m}$ while the $\sim$200 $\mu\textrm{m}$ active area thickness was produced by a deep diffusion doping profile, as in real HPK sensors. Each strip had a biasing electrode at zero potential as well as AC-coupled charge collecting contacts. The reverse bias voltage was provided by the backplane contact.
%
\section{Non-uniform three level defect model}\label{sec:4}
\paragraph{} As a first approach to find a defect model that will reproduce the expected surface properties after proton irradiation \cite{bib6}, \cite{bib7}, a shallow acceptor trap level, shown in table~\ref{tab:rad-dam-evl-3level}, is added to the proton model defects to provide increased negative space charge. Even though this results in good agreement with measured $R_{\textrm{\tiny int}}$ and $C_{\textrm{\tiny int}}$, the uniform distribution throughout the silicon bulk leads to unphysical suppression of the transient current and thus, to CCE values not matching to measurement. This suggests that a distribution in a limited region only is required for the shallow acceptor traps.
\begin{table}[tbp]
\caption{Three-level defect model included in the non-uniform 3-level model. $E_{\textnormal{\tiny C,V}}$ are the conduction and valence band energies, $\sigma$$_{\textnormal{\tiny e,h}}$ are the electron and hole capture cross sections, $\eta$ is the current introduction rate and $\Phi$ is the fluence.
}
\label{tab:rad-dam-evl-3level}
\smallskip
\centering
\begin{tabular}{lclclclclclc}
    \hline
    \bf{Defect} & \bf{Energy} \textnormal{[eV]} & \bf{$\sigma$$_{\textnormal{\tiny e}}$} \textnormal{[cm$^{2}$]} & \bf{$\sigma$$_{\textnormal{\tiny h}}$} \textnormal{[cm$^{2}$]} & \bf{$\eta$}
\textnormal{[cm$^{-1}$]} & \bf{Concentration} \textnormal{[cm$^{-3}$]}\\
    \hline
    Acceptor & $E_{\textnormal{c}}-0.525$ & 10$^{-14}$ & 10$^{-14}$ & $-$ & 1.189$\times$$\Phi$+6.454$\times$10$^{13}$\\
    Donor & $E_{\textnormal{v}}+0.48$ & 10$^{-14}$ & 10$^{-14}$ & $-$ & 5.598$\times$$\Phi$-3.959$\times$10$^{14}$\\
    Shallow acceptor & $E_{\textnormal{c}}-0.40$ & 8$\times$10$^{-15}$ & 2$\times$10$^{-14}$ & 40 & 40$\times$$\Phi$\\
    \hline
\end{tabular}
\end{table}
%
\subsection{From uniform to non-uniform defect concentration}\label{sec:4a}
\begin{figure}[tbp] 
\centering
\includegraphics[width=.5\textwidth]{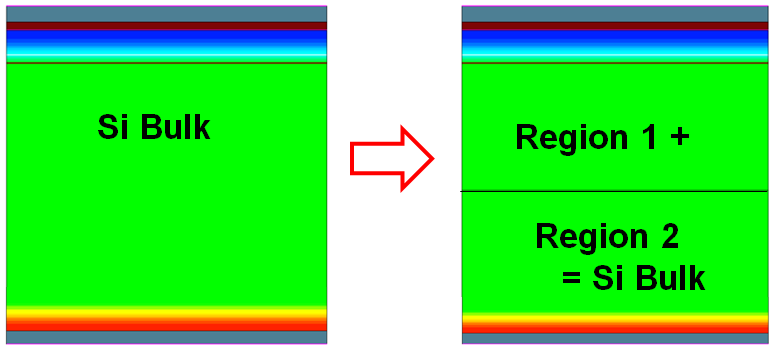}
\caption{Silicon bulk constructed by two methods (not to scale): (left) solid 300 $\mu\textrm{m}$ bulk, (right) 5 $\mu\textrm{m}$ + 295 $\mu\textrm{m}$ slabs, corresponding to a bulk thickness of 300 $\mu\textrm{m}$.}
\label{fig:2}
\end{figure}

\begin{figure}[tbp] 
\centering
\includegraphics[width=.25\textwidth]{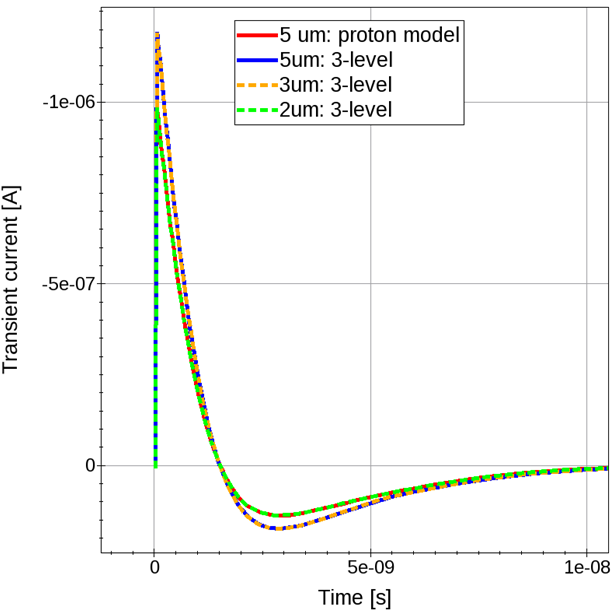}
\includegraphics[width=.36\textwidth]{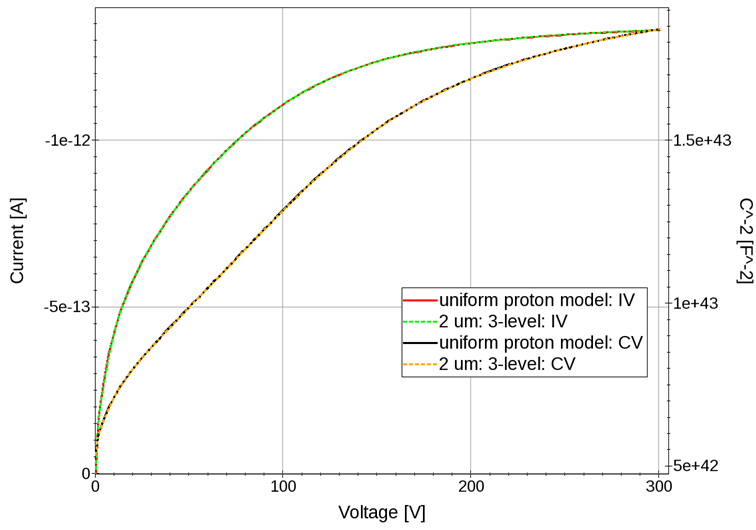}
\includegraphics[width=.25\textwidth]{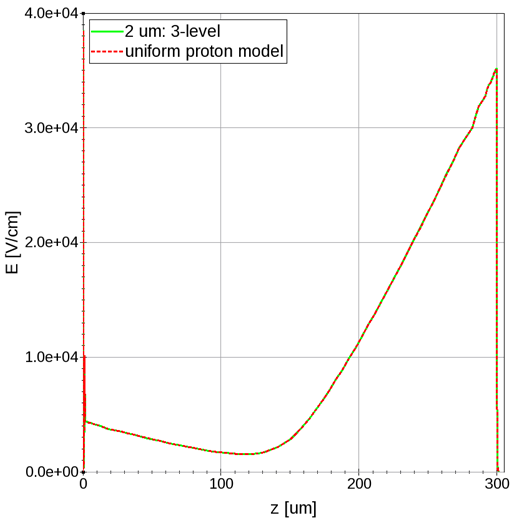}
\caption{(left) Variation of the thickness of the region 1 in figure~\protect\ref{fig:2} to find a transient current shape that matches the one in the proton model (red curve: proton model in both regions) in n-type diode structure at $\Phi_{\textrm{\tiny eq}}=5\times10^{14}$ $\textrm{cm}^{-2}$. (center) Corresponding $CV$, $IV$ and (right) $E$($z$) curves for 3-level model within 2 $\mu\textrm{m}$ depth.}
\label{fig:3}
\end{figure}

\paragraph{} Presented in figure~\ref{fig:2} are two methods to construct a silicon bulk for the simulation, namely a solid substrate and a combination of two slabs. Separate regions give the opportunity to introduce different defect concentrations into particular parts of the Si bulk, which would not be possible with a bulk constructed of only one region. When the change to the two region bulk structure was made, it was checked that the interface between the regions did not affect any of the simulated electrical properties of the device. Next the 3-level model in table~\ref{tab:rad-dam-evl-3level} is introduced to the region 1, close to the detector surface and the original proton model to the region 2 that forms the rest of the silicon bulk. Then the region 1 thickness is varied to preserve the bulk properties produced by the proton model. The left side of figure~\ref{fig:3} shows the thickness variation process for finding the transient signal shape equal to the proton model. 3-level model within 2 $\mu\textrm{m}$ depth produces matching transient current, leakage current-voltage ($IV$) and capacitance-voltage ($CV$) curves, as well as electric field distribution $E$($z$) (presented also in figure~\ref{fig:3}) in the silicon bulk. 

The implementation of the non-uniform defect concentration in the silicon bulk can be justified by the experimentally observed non-uniformity of the oxide concentration, that can indicate the existance of a higher number of traps and silicon impurities near the surface of an irradiated detector \cite{bib8}.  
%
\subsection{Surface properties simulation results}\label{sec:4b}
\begin{figure}[tbp] 
\centering
\includegraphics[width=.5\textwidth]{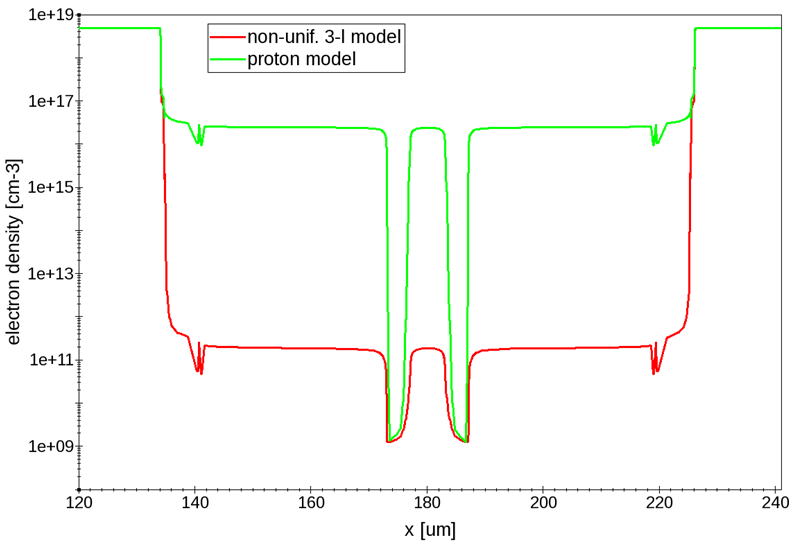}
\caption{Electron density for $\Phi_{\textrm{\tiny eq}}=1.5\times10^{15}$ $\textrm{cm}^{-2}$ and $Q_{\textrm{\tiny f}}=1.2\times10^{12}$ $\textrm{cm}^{-2}$ in a 5-strip 200P, pitch = 120 $\mu\textrm{m}$ detector structure. The cut was made at 50 nm below oxide between the n$^+$ strips. Positions of the double p-stops can be observed in the center of the plot, where both concentrations reach their respective minima.}
\label{fig:4}
\end{figure}

\paragraph{} When the 3-level model is applied to the region 1 with 2 $\mu\textrm{m}$ thickness in a 200P (pitch = 120 $\mu\textrm{m}$) strip-sensor, the expected high $R_{\textrm{\tiny int}}$ is reached already at $V=0$ for $\Phi_{\textrm{\tiny eq}}=5\times10^{14}$ $\textrm{cm}^{-2}$ and $Q_{\textrm{\tiny f}}=5\times10^{11}$ $\textrm{cm}^{-2}$. Also the geometrical value of $C_{\textrm{\tiny int}}$ up to $Q_{\textrm{\tiny f}}=1.5\times10^{12}$ $\textrm{cm}^{-2}$ at $\Phi_{\textrm{\tiny eq}}=1.5\times10^{15}$ $\textrm{cm}^{-2}$ is reproduced. The effect of shallow acceptor traps results in $\mathcal{O}(5)$ difference in electron density at the interstrip region, illustrated in figure~\ref{fig:4}. 

\begin{figure}[tbp] 
\centering
\includegraphics[width=.25\textwidth]{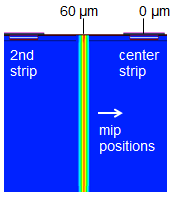}
\includegraphics[width=.6\textwidth]{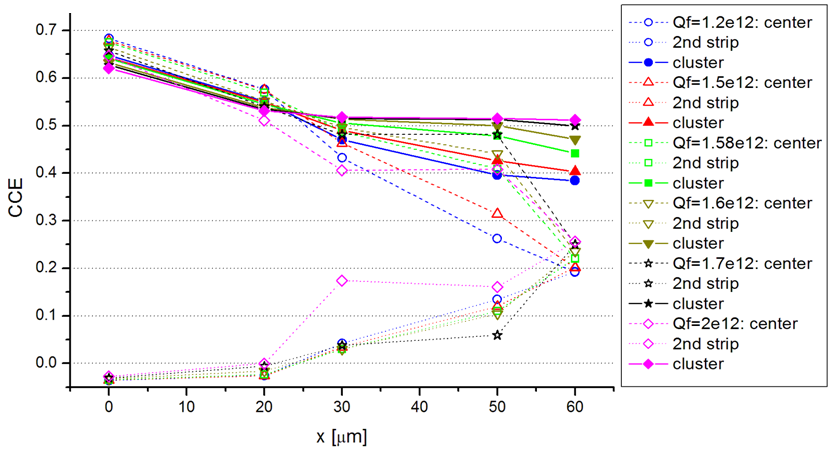}
\caption{(left) Principle of CCE($x$) simulation in a 5-strip 200P, pitch = 120 $\mu\textrm{m}$ structure. The minimum ionizing particle (MIP) injection position is varied towards the center strip. (right) Non-uniform 3-level defect model produced CCE($x$) in the same structure with p-stop doping $N_{\textrm{\tiny p}}=1\times10^{16}$ $\textrm{cm}^{-3}$. The cluster charge is calculated as the sum of signals collected at the two strips. Strips remain isolated for all values of $Q_{\textrm{\tiny f}}$ in the scan.}
\label{fig:5}
\end{figure}

The average cluster CCE loss between the strips measured with the Silicon Beam Telescope (SiBT) \cite{bib9} was determined to be $30\pm2\%$ \cite{bib10} (comparison of the collected charge after charge injection in the middle of the pitch and at the center of the strip) in 200 $\mu\textrm{m}$ active thickness float zone and magnetic-Czochralski sensors with p-stop and p-spray isolations (FZ200P/Y and MCz200P/Y, respectively) at $\Phi_{\textrm{\tiny eq}}=(1.4\pm0.1)\times10^{15}$ $\textrm{cm}^{-2}$. To reproduce this with simulation a procedure presented on the left side of figure~\ref{fig:5} was applied. On the right side of figure~\ref{fig:5} the collected charge of an irradiated detector is a measure of efficiency relative to the non-irradiated detector. The CCE loss is defined as the ratio of the difference in the collected cluster charge when the charge injection is made at the center of the strip and in the middle of the pitch, to the collected cluster charge when the injection position is at the center of the strip. For the cluster charge the sum of the collected charges at the two strips closest to the position of the charge injection was used.
When $Q_{\textrm{\tiny f}}$ is used as an iteration parameter, the simulation produces $\sim$31\% cluster CCE loss at $Q_{\textrm{\tiny f}}=1.58\times10^{12}$ $\textrm{cm}^{-2}$ (solid green curve on the right side of figure~\ref{fig:5}). Furthermore, the strips remain isolated throughout the scanned $Q_{\textrm{\tiny f}}$ range up to the high value $2\times10^{12}$ $\textrm{cm}^{-2}$. Shorted strips would result in equal collected charge at the centermost and its neighbour strip at all investigated charge injection positions.

%
\section{Conclusions}\label{sec:5}
\paragraph{} By combining a non-uniform concentration of 3-level defect model with the proton model in Synopsys Sentaurus simulations, it is possible to reproduce both bulk and surface damage dependent properties of irradiated silicon strip detectors at fluences up to $\Phi_{\textrm{\tiny eq}}=1.5\times10^{15}$ $\textrm{cm}^{-2}$. Also the measured position dependency of CCE was succesfully reproduced at a fixed fluence of $\Phi_{\textrm{\tiny eq}}=1.5\times10^{15}$ $\textrm{cm}^{-2}$. 
The CCE loss results showed here are strictly tuned for the aforementioned fluence. However, we know that the three levels are necessary also for low fluence although the parametrization used here need to be studied further and tuned in detail.

The effect of the non-uniform 3-level model to the CCE($x$) leads to the interpretation that the acceptor traps remove both inversion layer and signal electrons, i.e. the better the radiation damage induced strip isolation, the higher the CCE loss between the strips. On the other hand, if the shallow acceptor concentration is kept constant and $Q_{\textrm{\tiny f}}$ is increased, more traps are filled leading to increased charge sharing and decreased CCE loss between the strips. 

Similarly to Synopsys Sentaurus non-uniform 3-level model presented in this paper, the Silvaco Atlas 5-trap model is able to suppress the electron accumulation layer formation also at higher values of $Q_{\textrm{\tiny f}}$ \cite{bib11}, \cite{bib12}, \cite{bib13}. Thus, the qualitative match between the two defect models is apparent.




\begin{thebibliography}{9}


\bibitem{bib1}
R. Eber,
\emph{Investigations of new sensor designs and development of an effective radiation damage model for the simulation of highly irradiated silicon particle detectors},
\href{http://ekp-invenio.physik.uni-karlsruhe.de/record/48328/files/EKP-2014-00012.pdf} 
{\emph{PhD thesis}, Karlsruhe Institute of Technology, 2013, {\bf IEKP-KA/2013-27}}.

\bibitem{bib2}
RD50 collaboration, \emph{RD50 Status Report 2009/2010 - Radiation hard semiconductor devices for very high luminosity colliders},
\href{http://cds.cern.ch/record/1455062/files/LHCC-SR-004.pdf?version=1}
{CERN-LHCC-2012-010, LHCC-SR-004}.

\bibitem{bib3}
J. Zhang, E. Fretwurst, R. Klanner, I. Pintilie, J. Schwandt, M. Turcato, \emph{Investigation of X-ray induced radiation damage at the Si-SiO2 interface of silicon sensors for the European XFEL},
\jinst{7}{2012}{C12012}
[\href{http://arxiv.org/abs/1210.0427}
{\tt {arXiv:1210.0427}}].

\bibitem{bib4}
V. Eremin, E. Verbitskaya, A. Zabrodskii, Z. Li, J. H{\"{a}}rk{\"{o}}nen,
\emph{Avalanche effect in Si heavily irradiated detectors:
                  Physical model and perspectives for application},
\href{http://dx.doi.org/10.1016/j.nima.2011.05.002}
{\emph{Nucl. Instrum. $\&$ Meth. A} {\bf 658} (2011) 145 - 151}.

\bibitem{bib5}
K.-H. Hoffmann et al.,
\emph{Campaign to identify the future CMS tracker baseline},
\href{http://dx.doi.org/10.1016/j.nima.2011.05.028}
{\emph{Nucl. Instrum. $\&$ Meth. A} {\bf 658} (2011) 30 - 35}.

\bibitem{bib6}
A. Dierlamm, \emph{Characterisation of silicon sensor materials and designs for the CMS tracker upgrade}, in
\emph{The 21st International Workshop on Vertex Detectors},
\pos{PoS(Vertex 2012)016}.

\bibitem{bib7}
Y. Unno et al.,
\emph{p-bulk silicon microstrip sensors and irradiation},
\href{http://dx.doi.org/10.1016/j.nima.2007.05.256}
{\emph{Nucl. Instrum. $\&$ Meth. A} {\bf 579} (2007) 614 - 622}.

\bibitem{bib8}
A. Junkes,
\emph{Influence of radiation induced defect clusters on silicon particle detectors},
\href{http://www.iexp.uni-hamburg.de/groups/pd/sites/default/files/Alexandra_Junkes.pdf} 
{\emph{PhD thesis}, DESY, Hamburg, 2011}.

\bibitem{bib9}
C. Eklund et al.,
\emph{Silicon beam telescope for CMS detector tests},
\href{http://ac.els-cdn.com/S0168900299002107/1-s2.0-S0168900299002107-main.pdf?_tid=391ddf70-4fc9-11e4-b23c-00000aab0f01&acdnat=1412868819_6bf256e61c7fc9e5953430c51ab7fc15}
{\emph{Nucl. Instrum. $\&$ Meth. A} {\bf 430} (1999) 321 - 332}.

\bibitem{bib10}
T. M{\"{a}}enp{\"{a}}{\"{a}} et al., \emph{Performance of different silicon materials for the upgraded CMS tracker}, in proceedings of
\emph{RD13 11th International Conference on Large Scale Applications and Radiation Hardness of Semiconductor Detectors}, 2013
\pos{PoS(RD13)015}.

\bibitem{bib11}
R. Dalal, A. Bhardwaj, K. Ranjan, M. Moll, A. Elliott-Peisert, \emph{Combined effect of bulk and surface damage on strip 
		insulation properties of proton irradiated n$^+$-p silicon strip sensors}, 
		\jinst{9}{2014}{P04007}.

\bibitem{bib12}
T. Eichhorn et al., \emph{Simulations of inter-strip capacitance and resistance for the design of the CMS tracker upgrade}, in
\emph{Technology and Instrumentation in Particle Physics 2014}, 
\pos{PoS(TIPP2014)279}.

\bibitem{bib13}
R. Dalal et al., \emph{Development of radiation damage models for irradiated silicon sensors using TCAD tools}, in
\emph{Technology and Instrumentation in Particle Physics 2014}, 
\pos{PoS(TIPP2014)276}.

\end{thebibliography}
\end{document}